\documentclass[twocolumn,nodate,showkeys,superscriptaddress,preprintnumbers,amsmath,amssymb,aps,prd]{revtex4}
\usepackage{amsfonts}
\usepackage{latexsym}
\usepackage{mathptmx}
\usepackage{amsmath,amsfonts,latexsym,amssymb}
\usepackage[mathscr]{eucal}
\usepackage{mathrsfs}
\usepackage[bf,small]{caption2}
\usepackage{float}
\usepackage{graphicx}
\usepackage{amsmath}
\usepackage{amssymb}
\def\lb{\lambda}

\def\sq{\sqrt{q}}
\def\f{\frac}

\def\ka{\kappa}

\allowdisplaybreaks

\begin{document}
\title{ New formulation of Horava-Lifshitz quantum gravity as a master constraint theory}
\author{Chopin Soo}
\email{cpsoo@mail.ncku.edu.tw}
\author{Jinsong Yang}
\email{yangksong@gmail.com}
\affiliation{Department of Physics, National Cheng Kung University, Tainan, Taiwan}
\author{Hoi-Lai Yu}
\email{hlyu@phys.sinica.edu.tw}
\affiliation{Institute of Physics, Academia Sinica, Nankang, Taipei, Taiwan}

\begin{abstract}
Both projectable and non-projectable versions of Horava-Lifshitz gravity face serious challenges.
In the non-projectable version, the constraint algebra is seemingly inconsistent. The projectable version lacks a local Hamiltonian constraint, thus allowing for an extra scalar mode which can be problematic.
A new formulation of non-projectable Horava-Lifshitz gravity, naturally {\it realized as a representation of the master constraint algebra} studied by loop quantum gravity researchers, is presented.
 This yields a consistent canonical theory with first class constraints.  It captures the essence of Horava-Lifshitz gravity in retaining only spatial diffeomorphisms (instead of full space-time covariance)
 as the physically relevant non-trivial gauge symmetry; at the same time the local Hamiltonian constraint needed to eliminate the extra mode is equivalently enforced by the master constraint.

\end{abstract}
\maketitle

\section{Introduction}
 Horava's proposal \cite{Horava} of an ultraviolet completion of general relativity has attracted much recent attention. The conceptual breakthrough in Horava's work is the realization that a key obstacle to the viability of perturbative quantum gravity as a renormalizable field theory lies in the deep conflict between unitarity and space-time general covariance: the renormalizability of general relativity can be improved and achieved through the introduction of higher derivative terms, but space-time covariance requires higher time as well as spatial derivatives of the same order, thus compromising the stability and unitarity of the theory. The loss of unitarity is signaled by the occurrence
 of a ghost term in the resummed effective graviton propagator. Horava elects to keep unitarity but relinquishes full space-time covariance to retain only spatial diffeomorphism symmetry at the fundamental level, and seeks to recover general relativity at low curvatures.
 This bold procedure leads to the crucial decoupling of temporal from spatial contributions in the graviton propagator. There is, in our view, a related development, in loop quantum gravity, which is perhaps not as well known:
 the application of the master constraint program \cite{Thiemann,TD} to non-perturbative quantization of Einstein's theory. In the effort it has been fruitful to seek representations not of the Dirac algebra,
 but of the master constraint algebra which has the advantages of having structure constants rather than structure functions, and of spatial diffeomorphisms forming an ideal
 (thus allowing for the crucial decoupling of the equivalent quantum Hamiltonian constraint from spatial diffeomorphism generators).

 This work demonstrates that it is apposite, and also natural, to formulate Horava-Lifshitz theory as a representation of the master constraint algebra. Unlike the original non-projectable formulation of Horava-Lifshitz theory,
 our construction yields a consistent canonical theory with first class constraints. Observables of the resultant theory are also discussed.
 In addition our formulation realizes, in an explicit manner, the claim in Ref. \cite{Henneaux} that time-reparametrization symmetry of Horava-Lifshitz gravity is on-shell trivial.
The consistency conditions for general relativity with deformed supermetric are also discussed, and it is shown that  the special case covered
in Ref.\cite{Bellorin} can be consistent only with vanishing lapse function.

\section{Features of Horava-Lifshitz gravity with detailed balance}
We begin by recasting the original formulation of Horava-Lifshitz gravity \cite{Horava} in a compact form which highlights certain features and structures of the theory. In the canonical mould, the theory may be written as
\begin{equation}
\vspace*{-3mm}
S = \int \, \pi^{ij}{\dot q}_{ij} \,\,d^3x\, dt - \int\, (N H + N^iH_i)\, d^3x\, dt,
\end{equation}
wherein the super-Hamiltonian, with detailed balance \cite{Horava}, is
$H = \frac{\kappa^2}{2}\frac{G_{ijkl}}{\sqrt{q}}[\pi^{ij}\pi^{kl} +\frac{\delta W_{\rm T}}{\delta q_{ij}} \frac{\delta W_{\rm T}}{\delta q_{kl}}]$;
and $H_i = 2q_{ij}\nabla_k \pi^{kj} = 0$ is the super-momentum constraint. The space-time metric is of the Arnowitt-Deser-Misner (ADM) form \cite{ADM}
 $ds^{2}=-N^{2}(cdt)^{2}+q_{ij}(dx^{i}+N^{i}cdt)(dx^{j}+N^{j}cdt)$; and we assume there is no boundary for simplicity, although boundary terms can be added without affecting the theme of this work. The (inverse) DeWitt supermetric, with deformation parameter $\lambda$ which is allowed to deviate from unity, is $G_{ijkl} = \frac{1}{2}(q_{ik}q_{jl} +  q_{il}q_{jk}) - \frac{\lambda}{3\lambda -1} q_{ij}q_{kl}$. Dependent only on the 3-geometry, $W_{\rm T}$ is (up to 3rd order in spatial derivatives of the metric) the sum of a Chern-Simons action of the spatial affine connection and the spatial Einstein-Hilbert action with cosmological constant i.e.  $W_{\rm T} = W_{\rm CS} + W_{{\rm EH}\Lambda}$, with
 \begin{eqnarray}
 \vspace*{-3mm}
 W_{\rm CS} &=& \frac{1}{4{w}^2}\int{\tilde\epsilon}^{ikj}(\Gamma^l_{im} \partial_j\Gamma^m_{kl} +\frac{2}{3}\Gamma^l_{im}\Gamma^m_{jn}\Gamma^n_{kl})\,d^3x, \notag\\
 W_{{\rm EH}\Lambda}&=& \frac{\mu}{2}\int\,\sqrt{q}(R - 2\Lambda_W)\,d^3x;
\end{eqnarray}
and the Cotton tensor density can be expressed as $\tilde C^{ij} = {w}^2\frac{\delta W_{\rm CS}}{\delta q_{ij}}.$
Precisely because of detailed balance the Hamiltonian constraint can be succinctly rewritten as \footnote{After our initial discussion in Ref.\cite{SHY}, this particular form of the Hamiltonian constraint has also been investigated in Ref.\cite{Horava2}.}
\begin{equation}
\vspace*{-3mm}
H = \frac{\kappa^2}{2\sqrt{q}}G_{ijkl} Q_+^{ij}Q_-^{kl} =0, \label{H}
\end{equation}
with $Q^{ij}_\pm :=\pi^{ij} \pm i\frac{\delta W_{\rm T}}{\delta q_{ij}}$. As quantum operators,  ${\hat Q}_\pm $ take on the interesting form  ${\hat Q}^{ij}_\pm = e^{\pm \frac{W_{\rm T}}{\hbar}}{\hat\pi}^{ij}e^{\mp \frac{W_{\rm T}}{\hbar}}$. Note also that $Q_\pm$ are hermitian conjugates of each other if $W_{\rm T}$ is hermitian (classically real), and are separately hermitian if $W_{\rm T}$ is purely anti-hermitian (classically pure imaginary).
In both cases, the classical expression for $H$ remains real  \footnote{
The classical Hamiltonian constraint reduces to the algebraic equation $\frac{1}{2}{\rm Tr}(Q_+Q_- + Q_-Q_+) -\frac{\lambda}{3\lambda -1}{\rm Tr}(Q_+){\rm Tr}(Q_-) = 0$,
wherein the trace is taken with the indices of $Q^{ij}_\pm$ raised and lowered by the spatial metric.}.

An observation on the values of the coupling constants is also apposite here:
the superspace metric \cite{DeWitt}, $\delta S^2 \equiv G^{ijkl}\delta q_{ij}\delta q_{kl}$, has signature $({\rm sgn}[\frac{1}{3} -\lambda], +,+,+,+,+)$.
So the corresponding Wheeler-DeWitt equation \cite{DeWitt,Wheeler} comes equipped with an `intrinsic time'  \cite{DeWitt} for $\lambda > \frac{1}{3}$. Unlike the space-time covariant Einstein-Hilbert theory, deformation of $\lambda$ from unity should be allowed as it does not violate the 3-dim. diffeomorphism symmetry of the theory. Moreover, $\lambda$ is expected to flow as a renormalization parameter. Intriguingly, the emergent speed of light $c =\frac{\kappa^2\mu}{4}\sqrt{\frac{\Lambda_W}{1-3\lambda}}$, the cosmological constant $\frac{3}{2}\Lambda_W$,
and Newton's gravitational constant $G = \frac{\kappa^2c^3}{32\pi}$ can all be phenomenologically positive for $\lambda >\frac{1}{3}$  only if $\mu$ is pure imaginary and $\kappa$
is real. Then $H$ and the action is real only if $w^2$ is pure imaginary. This set of values renders $W_{\rm T}$ to be pure imaginary, and thus $Q_\pm$ become individually hermitian.

The supermetric and $Q_\pm$ do not commute among themselves, so whether $\Psi_{Q_\pm} = Ie^{\pm \frac{W_{\rm T}}{\hbar}}$,
which are annihilated by ${\hat Q}^{ij}_\pm$ (if $I$ satisfies $\frac{\delta I}{\delta q_{ij}} =0$), qualify as exact solutions depends on ordering ambiguities in $H$;
but it should be noted that such states (with slowly varying $I$) are nevertheless semi-classical; and a pure imaginary $W_{\rm T}$ leads to real $\pi^{ij} = \mp i\frac{\delta W_{\rm T}}{\delta q_{ij}}$
solving the Hamilton-Jacobi equation with $\pm iW_{\rm T}$ as Hamilton functions \footnote{$W_{\rm CS}$ (hence $W_{\rm T}$) is not invariant under large gauge transformations; so $\Psi_{Q_\pm}$,
 like $\theta$-angle states, will then acquire an additional phase factor. For the case of real $W_{\rm T}$, $\Psi_{Q_\pm}$ are not large
gauge-invariant even up to a phase, and they should, on this basis alone, be disqualified as physical states unless $I$ can compensate by producing an inverse factor \cite{soo}.}.
The form of the Hamiltonian with $Q_\pm$ gives a new and interesting perspective, not just on Horava-Lifshitz gravity (which corresponds to a theory with up to 3rd order spatial derivatives of the metric in $W_{\rm T}$); but also on the whole class of related theories which can be obtained by adjusting 3-geometry terms in $W_{\rm T}$. However, it is crucial that the constraint algebra must be consistent before any such theory is viable. Both the projectable and non-projectable version of the theory face serious challenges.

\section{Inconsistency of non-projectable Horava-Lifshitz gravity}
In the projectable version, with the lapse function, $N(t)$, dependent only on time, the theory has an integrated (rather than local) constraint $\int\, H\, d^3x =0$. With $H$ being a tensor density of weight one,
this integrated constraint commutes with itself, and also with the super-momentum constraint which generates spatial diffeomorphisms.
However, the absence of a local constraint $H({x})=0$ implies the theory has an additional unrestricted degree of freedom.
This extra scalar graviton mode is confirmed by the explicit analyzes \cite{Horava,Mukhohyama,others,Sasaki}, and the projectable version is pathological (and phenomenologically problematic) in having a perturbed Hamiltonian
which can be unbounded below. In the Einstein-Hilbert theory the would-be pathological scalar mode is eliminated precisely by the local Hamiltonian constraint, and in the weak field limit gravitons have but two polarizations.
Non-projectable Horava-Lifshitz gravity with space-time dependent lapse function $N(x)$ is also problematic. $\pi_N(x) =0 $ does lead to the secondary local constraint $H(x)=0$.
But the constraint algebra with the full $W_{\rm T}$ of Horava-Lifshitz gravity suffers from serious problems, whereas the Dirac algebra is obtained for covariant 4-dimensional Einstein-Hilbert action with cosmological constant.
In fact for Horava-Lifshitz gravity, $\{H(x),H[N]\} = (\vartriangle+\, \omega)N(x)$, wherein $\vartriangle$ contains spatial derivatives acting on $N$ and $\omega$ does not
(for explicit expressions of the Poisson bracket the reader may consult Refs. \cite{Henneaux,Li-Pang}), and smeared constraints are denoted by square brackets i.e. $H[N] :=\int NH d^3x$.
Furthermore $(\vartriangle+\,\omega)$ can have zero modes; and the only consistent solution is $N=0$ \cite{Henneaux}.
Strictly speaking, $N=0$ considered as a special case of a gauge-fixing condition $N=f$ results in $\{N({x}) -f, \pi_N({y})\} =\delta({x}-{y})$ with non-vanishing determinant.
So the condition does give a formally `consistent' system with two second class constraints, $\pi_N =N = 0 $, in addition to ${\pi}_{N^i}=H_i= H=0$ which are stable under evolution {\it provided} $N=0$.
But this formal consistency, achieved at the cost of vanishing $N$, is of dubious value (since any constraint can be made stable with $N=0$; furthermore the ADM space-time metric is degenerate for vanishing lapse function).
At the very least the situation demands a more physical explanation. On the other hand, Dirac's algorithm for the analysis of constrained canonical systems \cite{Dirac} should reveal the true gauge symmetries of the theory.
For Horava-Lifshitz gravity, the requirement of vanishing lapse function from the algorithm seems to signal that {\it only} 3-dim. spatial diffeomorphism symmetry is physically relevant.
Such theories cannot obey the Dirac algebra which is the hallmark of space-time covariance and the embeddability of hypersurface deformations, and from which Einstein's geometrodynamics can be uniquely recovered \cite{HKT,Teitelboim}.

There can be interesting modifications to the Dirac algebra in theories without full space-time covariance. For example, in the extreme limit of $W_{\rm T} = 0$ corresponding to ultra-local gravity with
$H=\f{2\ka'}{\sq}G_{ijkl}\pi^{ij}\pi^{kl}$, a strongly vanishing commutator, $\{H[N], H[M]\} =0$ (even for $\lambda \neq 1$) replaces the usual commutation relation in the Dirac algebra; with
 the special case of $\lambda =1$ already pointed out in Ref. \cite{Teitelboim}. When a scalar curvature term is added to the previous ultralocal theory in deformations of Einstein-Hilbert theory with $\lambda \neq 1$  and $H =(\f{2\ka'}{\sq}G_{ijkl}\pi^{ij}\pi^{kl}-\f{\sq}{2\ka'}\,R)$, stability of the primary constraints $\pi_N=\pi_{N^i}=0$ with respect to the Hamiltonian $H_{\rm{primary}} = \int d^3x\left(\Lambda  \pi_N+\Lambda^i\pi_{N^i}+NH+N^iH_i\right)$ results in $H=H_i=0$. Preservation of these secondary constraints under evolution leads to
\begin{align}
\vspace*{-3mm}
&\{H[M], H_{\rm{primary}}\}= -H[{\cal L}_{\vec N}M] +
H_i[(M\nabla^iN-N\nabla^iM)]\notag\\
&-\f{2(1-\lb)}{3\lb-1}\int (M\nabla^iN-N\nabla^iM)\nabla_i\pi\, d^3x\,.
\end{align} For general relativity with $\lambda =1$ the constraints are already first class at this stage. With $\lambda \neq 1$  (the case of degenerate supermetric with  $\lambda =\frac{1}{3}$ has been addressed previously \cite{Pons} and will not be taken up here)
the consequent secondary constraint $Z_i :=\nabla_i\pi=0$ leads to
\begin{align}
\vspace*{-3mm}
&\{Z_i[{\xi^i}],Z_j[\chi^j]\}=Z_i\left[\frac{3}{2}(\chi^i\nabla_j\xi^j-\xi^i\nabla_j\chi^j)\right],\notag\\
&\{H_i[N^i],Z_i[\xi^i]\}=Z_i\left[{\cal L}_{\vec N}\xi^i\right],\notag\\
&\{Z_i[\xi^i],H[N]\}
  =Z_i\left[-\f{2\ka'}{(3\lb-1)\sq}N\pi\xi^i\right]-H\left[\f{3}{2}N\nabla_i\xi^i\right]\notag\\
  &-\f{1}{\ka'}\int
  d^3x\,\sq\,(\nabla_j\xi^j)W,
\end{align}
with $W:=\left[-\nabla^2 + R +\f{2\ka'^2\pi^2}{(3\lb-1)q}\right]N$.
Thus $W=0$ is required for stability of $Z_i :=\nabla_i\pi=0 \Leftrightarrow \pi = K(t)\sqrt{q}$.
The constraint $H=0$ allows us to write $R = \f{4\ka'^2}{q}\left({\overline{\pi}}_{ij}{\overline{\pi}}^{ij}-\f{1}{3(3\lb-1)}\,\pi^2\right)$,
wherein ${\overline{\pi}}^{ij} := \pi^{ij} -\f{1}{3}q^{ij}\pi$ is the traceless part of the momentum. Together with $\pi =K\sq$, the condition on $N$ then becomes
\begin{equation}
W=\left[-\nabla^2 +\f{4\ka'^2}{q}{\overline{\pi}}_{ij}{\overline{\pi}}^{ij} + \f{2\ka'^2 K^2}{3(3\lb-1)}\right]N=0.
\end{equation}
Since $-\nabla^2$ and  $\f{4\ka'^2}{q}{\overline{\pi}}_{ij}{\overline{\pi}}^{ij}$ are both positive semi-definite operators, $W=0 $ can have non-vanishing solution
\footnote{$N$ can be expanded in terms of eigenfunctions of the Hermitian operator $-\nabla^2 + \f{4\ka'^2}{q}{\overline{\pi}}_{ij}{\overline{\pi}}^{ij}$. Then  $N$ is non-trivial iff
$-\f{2\ka'^2 K^2}{3(3\lb-1)}$ coincides with at least one of its (positive semi-definite) eigenvalues. This can be achieved {\it only} for $\lambda < \frac{1}{3}$.} for $N$ only if $\lambda < \f{1}{3}$.
For $\lambda < \frac{1}{3}$, the resultant theory (counting the 6 conjugate pairs $(q_{ij}, \pi^{ij})$,  $H_i =0$ as first class and $H=0, \pi =K\sq$ as second class constraints) has $\frac{1}{2}[12 - 3(2)- 2(1)]=2$
degrees of freedom, but does not contain Einstein's theory ($\lambda =1$) as a special case.
For $\lambda > \frac{1}{3}$, we are lead to the fact $N=0$ is the only solution for $W=0$.
It also follows that the special case with vanishing  $\pi$ (or $K =0$) covered in Ref. \cite{Bellorin} can be consistent only for $N=0$.
In our more general context of allowing for $K \neq 0$, non-trivial $N$ exists for $\lambda < \frac{1}{3}$.
However, for $\lambda > \frac{1}{3}$ (and for non-projectable Horava gravity with local Hamiltonian constraint) only $N=0$ is allowed.

\section{Horava-Lifshitz gravity as a master constraint theory}
As structure {\it functions} are present in the commutator of two Hamiltonian constraints, the Dirac algebra is not the Lie algebra of 4-dimensional diffeomorphisms. But $H_i$ and $H$ constraints do generate 4-dimensional
diffeomorphisms on-shell(modulo the constraints and equations of motion). In a theory which possesses at the fundamental level only 3-dimensional diffeomorphisms as gauge symmetry, we expect the constraints to generate, {\it on-shell, only spatial diffeomorphisms}. There is a formulation which precisely achieves this goal, and which at the same time gives rise to a condition equivalent to the local constraint $H(x)=0$ -- thus eliminating the problematic extra scalar graviton mode. For Horava-Lifshitz gravity, simultaneous requirement of a {\it local} $H$ constraint and an involutive constraint algebra seems impossible without $N=0$. Our proposal is for theories with {\it only} spatial diffeomorphism invariance as the physical gauge symmetry, in particular Horava-Lifshitz gravity, to be formulated as representations of the master constraint algebra which has been studied by researchers in loop quantum gravity in their attempt to quantize Einstein's theory non-perturbatively \cite{Thiemann,TD}. The master constraint operator $\mathbf{M}$, which is tailored to be invariant under spatial diffeomorphisms (with  $H(x)$ being a scalar density of weight 1), is defined as $ \mathbf{M}:=\int_\Sigma\,\f{[H({x})]^2}{\sqrt{q({ x})}}\,d^3x$.
For any real valued $H(x)$, the integrand is positive-semi-definite; and the master constraint equation, $\mathbf{M}=0$, is mathematically equivalent to $H(x)=0$ everywhere on the Cauchy surface $\Sigma$.
Thus the master constraint equation replaces the infinite number of local restrictions ($H = 0$) by a single global restriction.
This equivalence has furthermore been demonstrated rigorously in the quantum context for various non-trivial models, including for quantum field theories \cite{Thiemann,TD}.
The upshot is a simple closed constraint algebra (with structure {\it constants}) which is first class. The master constraint algebra is just
\begin{align}
\vspace*{-3mm}
  &\{H_i[N^i],H_j[{N'}^j]\}=H_i[{\cal
  L}_{\vec{N}}{N'}^i],\notag\\
  &\{H_i[N^i],\mathbf{M}\}=0\,,  \quad
\left\{\mathbf{M}\,,\mathbf{M}\right\}=0.
\end{align}
The canonical action for Horava-Lifshitz gravity can then be consistently adopted as
\begin{equation}
\vspace*{-3mm}
S = \int\pi^{ij}{\dot q}_{ij} \,d^3xdt - \int\frac{N(t)}{\epsilon_o}\mathbf{M}dt  - \int\, N^iH_i\, d^3xdt,
\end{equation}
with $H$ of the form in Eq.\eqref{H}; $\epsilon_o$ has the physical dimension of energy density. Such theories consistently generate equations of motion which are (on-shell) equivalent to spatial diffeomorphisms since
\begin{align}
\vspace*{-3mm}
\{q_{ij}, \frac{N(t)}{\epsilon_o}\mathbf{M} + H_k[N^k]\}|_{\mathbf{M}=0
\Leftrightarrow H=0}&\approx  \{q_{ij}, H_k[N^k]\}\notag\\
 &={\cal L}_{\vec N}q_{ij},\label{comut}
 \end{align}
  (and similarly for $\pi^{ij}$). Our new formulation also realizes the claim in Ref. \cite{Henneaux} that time-reparametrization symmetry of Horava-Lifshitz gravity and freedom in the choice of $N(t)$ is
  on-shell trivial. We should remark that in general relativity with first class local constraints $H=H_i =0$, a Dirac observable, $O$, must commute with both $H_i$ and $H$.
  This is equivalent to the requirement $\{O, \{O,\mathbf{M}\}\}|_{\mathbf{M}=0} = 0$ \cite{Thiemann}. But for the theory at hand, one can read off from the action that weak observables $O$ should commute (weakly)
  with $\mathbf{M}$ and $H_i$; which (analogous to computations in Eq.\eqref{comut}) is equivalent to $\{O, H_i\} \approx 0.$
  This weaker criterion (instead of also requiring vanishing $\{O, \{O,\mathbf{M}\}\}|_{\mathbf{M}=0}$) of allowing for observables of 3-geometry on the constraint surface is physically reasonable
  as observables of the theory should be invariant only with respect to the local gauge symmetry of spatial diffeomorphisms.
  In such theories, two configurations differing by four-dimensional, rather than spatial, coordinate transformations can be physically inequivalent.
  It is to be noted our formulation of Horava-Lifshitz gravity {\it also requires the lapse function to depend only on $t$}.
In retrospect, `troubles' in the constraint algebra of Horava-Lifshitz gravity with {\it local} Hamiltonian constraint are to be expected of a canonical theory of 3-geometry which fundamentally possesses only spatial diffeomorphism
invariance. A new, and natural, canonical formulation of Horava-Lifshitz gravity as a representation of the master constraint algebra can be consistently constructed.
Moreover, the local Hamiltonian constraint  which is needed (as in Einstein's theory) to remove the problematic scalar graviton mode, is equivalently enforced by the master constraint.
Although the Legendre transformation to Lagrangian formulation can be performed, it may not be particularly useful given the lack of 4-dimensional general covariance, and will not be pursued here as the canonical Hamiltonian formulation is already {\it manifestly covariant} with respect to the full spatial diffeomorphism symmetry of the theory. It is also noteworthy that, rather than working directly with the quantum version of the intractable Dirac algebra, the loop quantum gravity community has instead found it fruitful to seek quantum representations of general relativity through the master constraint algebra \cite{Thiemann}. Horava-Lifshitz gravity is in fact an explicit, and highly non-trivial representation, of a master constraint theory with spatial diffeomorphism invariance.
In not satisfying the Dirac algebra with local Hamiltonian constraint, Horava-Lifshitz gravity is in fact more naturally associated with the master constraint theory than Einstein's general relativity.
The expectation (albeit at the level of perturbative quantum field theory) of the absence of negative norm ghosts is also encouraging. With a positive-definite norm, the quantum theory with
${\hat{\mathbf{M}}} := \int \widehat{\frac{H^\dagger H}{\sqrt q} }\, d^3x$ does lead to ${\hat{\mathbf{M}}}|\Psi\rangle =0 \Leftrightarrow {\hat H}|\Psi\rangle =0$.
This new formulation of Horava-Lifshitz theory offers the exciting perspective that the perturbative as well as non-perturbative aspects of a theory of quantum gravity may become accessible through both the
methodologies of perturbative renormalizable quantum field theories and non-perturbative quantum representations of the master constraint algebra.

\section*{Acknowledgments}
This work has been supported in part by the National Science Council
of Taiwan under Grant Nos. NSC98-2112-M-006-006-MY3,
98-2811-M-006-035, 99-2811-M-006-015, 97-2112-M-001-005-MY3, and the National Center
for Theoretical Sciences, Taiwan. Beneficial interactions and
discussions with J. Fernando Barbero G. are also gratefully
acknowledged.


\section*{References}
\begin{thebibliography}{00}

\bibitem{Horava}
P. Horava, Phys. Rev. D {\bf{79}},  084008 (2009).
\bibitem{Thiemann}
T. Thiemann, Class. Quantum Grav. {\bf{23}},  2211 (2006).
\bibitem{TD}
B. Dittrich and T. Thiemann, Class. Quantum Grav. {\bf{23}}, 1025 (2006);
Class. Quantum Grav. {\bf{23}}, 1067 (2006); Class. Quantum Grav. {\bf 23}, 1089 (2006);
Class. Quantum Grav. {\bf{23}}, 1043 (2006).
\bibitem{Henneaux}
M. Henneaux, A. Kleinschmidt and G. LucenaGomez, Phys. Rev. D {\bf 81}, 064002 (2010); [arXiv:1004.3769].
\bibitem{Bellorin}
J. Bellorin and A. Restuccia, [arXiv:1004.0055].
\bibitem{ADM}
R.~L.~Arnowitt, S.~Deser and C.~W.~Misner, Phys.\ Rev. {\bf 116}, 1322 (1959).
\bibitem{SHY}
C. Soo, J. Yang and H. L. Yu, [arXiv:1007.1563v1].
\bibitem{Horava2}
P. Horava and C. M. Melby-Thompson, {\it Phys. Rev.} D {\bf 82}, 064027 (2010).
\bibitem{DeWitt}
Bryce S. DeWitt, Phys. Rev. {\bf 160}, 1113 (1967).
\bibitem{Wheeler}
J. A. Wheeler, {\it Superspace and the nature of quantum geometrodynamics}, in Battelle Rencontres, edited by C. M.
DeWitt and J. A. Wheeler (W. A. Benjamin, New York, 1968).
\bibitem{soo}
 C. Soo, Class. Quantum Grav. {\bf 19}, 1051 (2002)).
\bibitem{Mukhohyama}
K. Koyama and F. Arroja, JHEP {\bf 1003}, 061 (2010).
\bibitem{others}
C. Charmousis, G. Niz, A. Padilla and P. Saffin, JHEP {\bf 0908}, 070 (2009);
D. Blas, O. Pujolas and S. Sibiryakov, JHEP {\bf 0910}, 029 (2009).
\bibitem{Sasaki}
J.-O. Gong, S. Koh and M. Sasaki, Phys. Rev. D {\bf 81}, 084053 (2010);
A. Cerioni and R. Brandenberger, [arXiv:1007.1006].
\bibitem{Li-Pang}
M. Li and Y. Pang, JHEP {\bf{0908}}, 015 (2009).
\bibitem{Dirac}
P. A. M. Dirac, {\it Lectures in Quantum Mechanics} (Yeshiva University Press, New York, 1964).
\bibitem{HKT}
S. Hojman, K. Kuchar and C. Teitelboim, Nature Phys. Sci. {\bf 245}, 97 (1973); Ann. Phys. {\bf 96}, 88 (1976).
\bibitem{Teitelboim}
C. Teitelboim, {\it The Hamiltonian structure of spacetime}, in General Relativity and Gravitation Vol. 1, edited by A. Held (Plenum, New York, 1980).
\bibitem{Pons}P. Horava, JHEP {\bf 0903}, 020 (2009); J. M. Pons and P. Talavera, Phys. Rev. D {\bf 82}, 044011 (2010).







\end{thebibliography}
\end{document}